\documentclass{article}
\usepackage{spconf,amsmath,graphicx}
\usepackage{mathtools}
\usepackage{multirow}
\usepackage{caption}
\usepackage{subcaption}
\usepackage{url}
\usepackage{bm}
\usepackage{booktabs}
\let\OLDthebibliography\thebibliography
\renewcommand\thebibliography[1]{
  \OLDthebibliography{#1}
  \setlength{\parskip}{0pt}
  \setlength{\itemsep}{0pt plus 0.5ex}
}

\title{Personalized Lightweight Text-to-Speech: Voice Cloning with Adaptive Structured Pruning}
\name{Sung-Feng Huang$^1$, Chia-ping Chen$^2$, Zhi-Sheng Chen$^2$, Yu-Pao Tsai$^2$, Hung-yi Lee$^1$}
\address{$^1$National Taiwan University, $^2$Intelligo Technology Inc.\\
\texttt{f06942045@ntu.edu.tw, ailsa.chen@intelli-go.com, cs.chen@intelli-go.com,}\\\texttt{yptsai@gmail.com, hungyilee@ntu.edu.tw}}
\begin{document}
\ninept
\maketitle
\begin{abstract}
Personalized TTS is an exciting and highly desired application that allows users to train their TTS voice using only a few recordings. However, TTS training typically requires many hours of recording and a large model, making it unsuitable for deployment on mobile devices. To overcome this limitation, related works typically require fine-tuning a pre-trained TTS model to preserve its ability to generate high-quality audio samples while adapting to the target speaker's voice. This process is commonly referred to as ``voice cloning.'' Although related works have achieved significant success in changing the TTS model's voice, they are still required to fine-tune from a large pre-trained model, resulting in a significant size for the voice-cloned model. In this paper, we propose applying trainable structured pruning to voice cloning. By training the structured pruning masks with voice-cloning data, we can produce a unique pruned model for each target speaker. Our experiments demonstrate that using learnable structured pruning, we can compress the model size to 7 times smaller while achieving comparable voice-cloning performance.
\end{abstract}
\section{Introduction}
\label{sec:intro}

End-to-end text-to-speech (TTS) is a well-researched topic, but customization is an area that has not been thoroughly explored. To train a high-quality single-speaker end-to-end TTS~\cite{wang2017tacotron,shen2018natural,li2019neural,ren2019fastspeech,ren2020fastspeech}, hours of single-speaker speech recordings~\cite{ljspeech17} and large, specially designed models are required. However, customizing TTS voice usually entails requesting users to record hours of speech and then spending days training a large model, which is not always practical. Additionally, the ultimate goal of personalized TTS is to deploy on mobile devices, which eliminates concerns about personal data upload or personalized TTS storage in cloud storage. Therefore, three aspects need improvement to build an ideal personalized TTS application: limited training data, faster training speed, and smaller model size. Since it is challenging to train a TTS from scratch with limited data in experience, related works often use transfer learning to ensure the TTS synthesizes high-quality speech. This process, transferring a trained TTS with limited recording data of an unseen speaker, is also referred to as "voice cloning." For example, \cite{arik2018neural,chen2018sample,Wang20adaptation,chen2021adaspeech,Song21M2VoC} pre-train a multi-speaker TTS, then fine-tune the speaker embedding and/or the TTS model with the few-shot target speaker recordings. Other works learn a speaker encoder with the multi-speaker TTS model and expect the speaker encoder to generalize to unseen speakers without fine-tuning~\cite{jia2018transfer,cooper2020zero,Yaniv18Voiceloop,Choi20Attentron,Wang20bilevel,Cai20verification,Chien21M2VoC}. Meta-TTS~\cite{huang2022meta} and Meta-Voice~\cite{liu2021meta} leverage meta-learning to speed up the fine-tuning procedure, enabling the meta-learned TTS model to adapt to the target speaker's voice with fewer fine-tuning steps while still producing high-quality audio.

In personalized TTS, reducing the model size is crucial to avoid slower computational speed, higher computational costs, and increased local storage. Although there are few related works about minimizing the model size of an end-to-end TTS, none of them focuses on voice-cloned TTS. LightSpeech~\cite{luo2021lightspeech} employs neural architecture search within a limited-parameter search space to discover improved architecture that can reduce the model size while still preserving its performance. On the other hand, \cite{lai2022interplay} prunes a trained end-to-end TTS model with an unstructured pruning method, which makes the model sparse by zeroing out a portion of each weight matrix. However, sparse matrix computation necessitates specialized hardware for acceleration, making it difficult to take advantage of the reduced model size to boost computational speed and lower computational costs. Both of these works are designed for single-speaker end-to-end TTS and require hours of recording data (e.g., LJSpeech~\cite{ljspeech17}) for the training process to ensure the synthesized audio quality, which is unsuitable for voice cloning tasks.

\begin{figure}
    \centering
    \begin{subfigure}[b]{.35\linewidth}
        \centering
        \includegraphics[width=\linewidth]{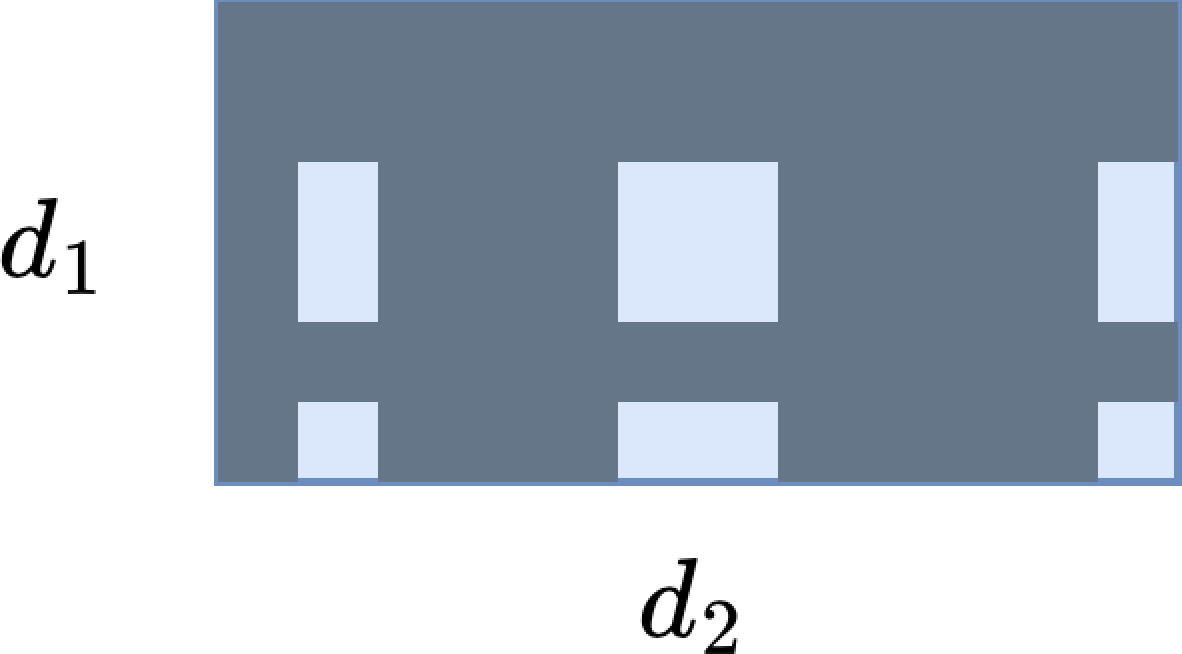}
        \caption{Structured pruning.}
        \label{subfig:structured}
    \end{subfigure}
    \hspace{30pt}
    \begin{subfigure}[b]{.35\linewidth}
        \centering
        \includegraphics[width=\linewidth]{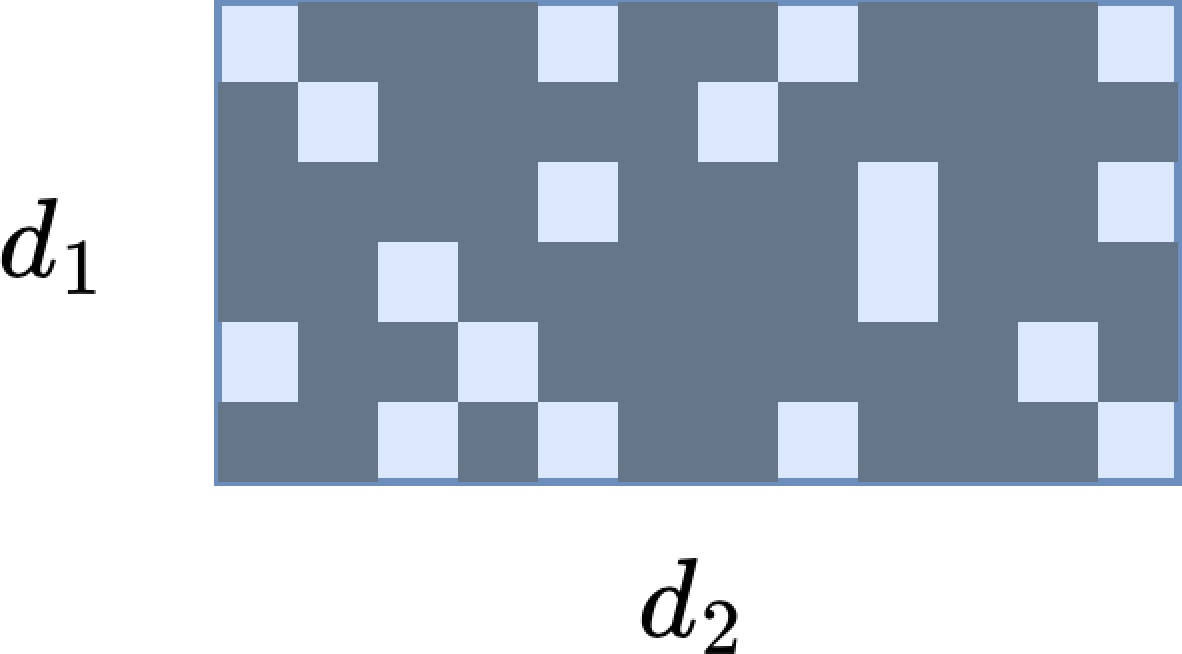}
        \caption{Unstructured pruning.}
        \label{subfig:unstructured}
    \end{subfigure}
    \caption{
        Illustration of how structured and unstructured pruning works.
        Fig.~\ref{subfig:structured} and \ref{subfig:unstructured} show how structured/unstructured pruning affect a $\mathcal{R}^{d_1 \times d_2}$ weight matrix, respectively.
        The gray regions are masked (pruned) while not the blue parts.
        For structured pruning, we can concatenate the kept parts into a smaller matrix, while we can't do the same to the unstructured pruned matrix.
    }
    \label{fig:structured-prune}
\end{figure}

This paper proposes utilizing a learnable structured pruning method for the voice cloning task. Unlike unstructured pruning, structured pruning eliminates channels (dimensions) of each weight matrix, resulting in a smaller weight matrix instead of a sparse matrix, thereby accelerating matrix computation and reducing computational costs, as demonstrated in Figure~\ref{fig:structured-prune}.
Additionally, whereas pruning methods commonly rely on pruning parameters based on criteria such as weight magnitude, we propose structured pruning with learnable masks to determine which channels to prune. By doing so, we can train a personalized pruning mask for each target speaker, resulting in a lightweight customized TTS model.
Furthermore, the pruning procedure can be utilized before, after, or in conjunction with the fine-tuning stage of the voice cloning task, which we will explore further in our experiments. To our knowledge, we are the first to employ adaptive structured pruning in the voice cloning task and the first to train a learnable pruning mask using few-shot data only (8-shot in our experiments, equivalent to approximately 24 seconds in total).

\section{Background}
\label{sec:background}
In this paper, we utilize FastSpeech 2~\cite{ren2020fastspeech} as the TTS model architecture. Further details regarding implementation and the speaker embedding lookup table's utilization to construct a multi-speaker FastSpeech 2 can be found in our prior work~\cite{huang2022meta}.

\subsection{Voice cloning}
Voice cloning involves creating a TTS model of the target speaker's voice with only a few-shot dataset. As mentioned in the introduction, training a TTS model from scratch with limited data may cause overfitting and low-quality audio generation. As a result, fine-tuning from a pre-trained multi-speaker TTS model is typically employed. Most existing works utilize multitask learning to pre-train the TTS~\cite{arik2018neural,chen2018sample,Wang20adaptation,chen2021adaspeech,Song21M2VoC}. Some other works use meta-learning~\cite{huang2022meta,liu2021meta} to expedite the fine-tuning process.

\subsection{Transformer blocks}
Our FastSpeech 2 model comprises an encoder, a decoder, a variance adaptor, an output linear layer after the decoder to generate Mel-spectrograms, and a post-net to add more details to the output through a residual connection. The encoder and decoder are built using stacks of Transformer blocks, whereas the variance adaptor and the post-net are composed of CNN layers. Each Transformer block includes a multi-head self-attention (MHA) layer and a feed-forward (FFN) layer.
We formulate a self-attention layer with input $X\in \mathcal{R}^{L \times d}$ as below:
\begin{equation}
    \mathrm{SelfAtt}(W_Q,W_K,W_V,X)=\mathrm{softmax}(\frac{XW_QW_K^\top X^\top}{\sqrt{d_k}})XW_V
    \label{eq:self-att}
\end{equation}
$L$ and $d$ represent the length and hidden dimension of $X$, respectively. $d_k$ denotes the hidden dimension of the self-attention layer, and $W_Q$, $W_K$, and $W_V \in \mathcal{R}^{d \times d_k}$ are the query, key, and value matrices, respectively.
Then an MHA layer with $N_h$ heads takes an input $X$ would output:
\begin{equation}
    \mathrm{MHA}(X) = \sum_{i=1}^{N_h}\mathrm{SelfAtt}(W_Q^{(i)},W_K^{(i)},W_V^{(i)},X)W_O^{(i)}
    \label{eq:mha}
\end{equation}
where $W_O\in \mathcal{R}^{d_k \times d}$ denotes the output matrix and $W_Q^{(i)}$, $W_K^{(i)}$, $W_V^{(i)}$, $W_O^{(i)}$ represent the matrices of each head.

Also, we can formulate a Feed-forward layer as below, which includes an up-projection and a down-projection layer:
\begin{equation}
    \mathrm{FFN}(X)=\mathrm{ReLU}(XW_U)W_D
    \label{eq:ffn}
\end{equation}
where $W_U\in \mathcal{R}^{d \times d_f}$ and $W_D \in \mathcal{R}^{d_f\times d}$ represent the up-projection and down-projection layer respectively.

The output of a Transformer block can then be formulated as below, where $LN$ indicates layer norm:
\begin{equation}
    \begin{gathered}
        \label{eq:transformer-block}
        X'=\mathrm{LN}(\mathrm{MHA}(X)+X)\\
        \mathrm{TransformerBlock}(X)=\mathrm{LN}(\mathrm{FFN}(X')+X').
    \end{gathered}
\end{equation}

\subsection{Structured pruning}
\begin{figure}
    \centering
    \includegraphics[width=.7\linewidth]{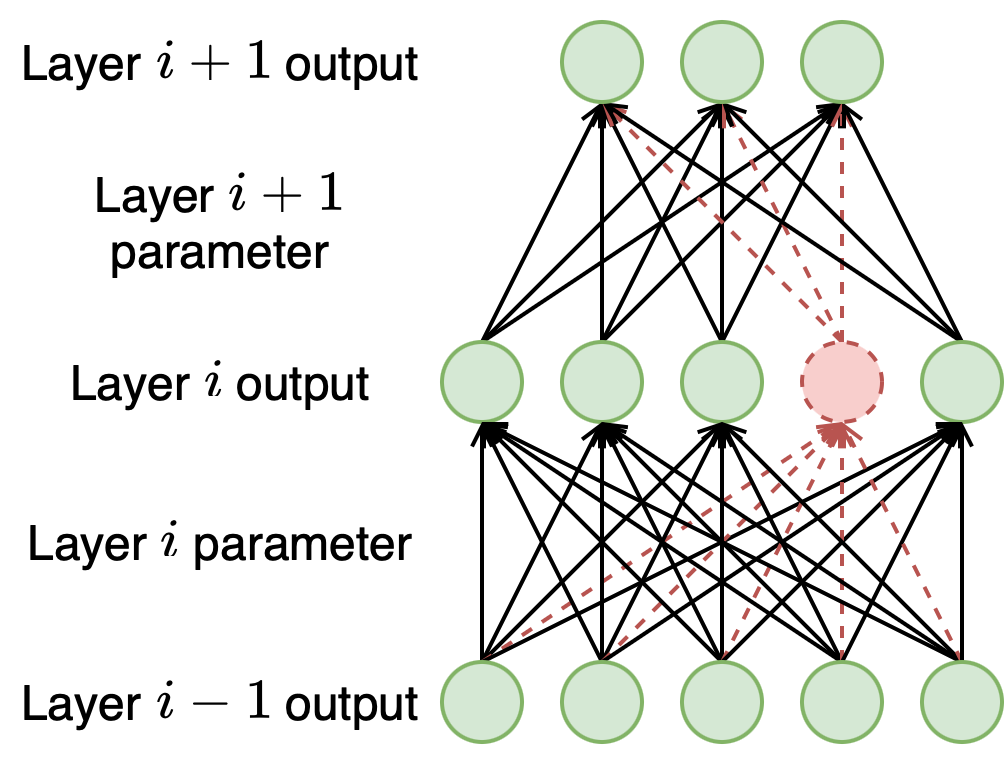}
    \caption{
        Illustration of what happens when we prune the red neuron. The red dashed arrows are the parameters removed together with the red neuron, while the black arrows are the parameters kept. The output dimension of layer $i$ and the input dimension of layer $i+1$ are therefore reduced by 1.
    }
    \label{fig:prune}
\end{figure}

Unlike unstructured pruning, which selects individual model parameters (i.e., elements in weight matrices) to discard, structured pruning identifies specific neurons in each layer's output to eliminate. This method removes the dropped neurons and their corresponding channels in adjacent parameters, as illustrated in Figure~\ref{fig:prune}.

\subsection{Pruning with $L_0$ regularization}
\label{subsec:l0}
Most model pruning techniques determine their binary pruning mask based on some criteria, such as the magnitude of the parameters. However, these criteria are not suitable for personalizing pruning masks for each target speaker. To address this issue, \cite{louizos2017learning} proposes training the binary pruning mask by adding a regularization term during training, specifically the $L_0$ norm of the binary pruning masks. Since discrete binary masks are not differentiable, \cite{louizos2017learning} utilizes the hard-concrete distribution to relax the binary masks into continuous and make them trainable.
As a result, the regularization term becomes the $L_1$ norm of the masks.

We could sample a learnable mask $\mathbf{z}$ from the hard-concrete distribution as follows:
\begin{equation}
\begin{aligned}
    \label{eq:mask}
    \mathbf{u}&\sim U(0,1)\\
    \mathbf{s}&=\mathrm{Sigmoid}\left(\frac{\log \mathbf{u} - \log \mathbf{(1-u)} + \log \mathbf{\alpha}}{\beta}\right)\\
    \mathbf{z}&=\mathrm{min}\left(1, \mathrm{max}\left(0, \gamma + \mathbf{s}\left(\eta-\gamma\right)\right)\right)
\end{aligned}
\end{equation}
Where we denote $\mathbf{u}$ as a random variable sampled from a uniform distribution $U(0,1)$. Hyperparameters $\gamma\leq 0$ and $\eta\geq 1$ are used to stretch the output interval of the Sigmoid function from $(0,1)$ to $(\gamma,\eta)$, while hyperparameter $\beta$ controls the steepness of the function. The main learnable masking parameter is $\log\mathbf{\alpha}$, which represents the logit of the Bernoulli distribution where $\mathbf{z}$ is sampled from.
To distribute the output $\mathbf{s}$ in the interval $(0,1)$ with a probability derived from a relaxed continuous version of $\mathrm{Bernoulli}(\mathrm{Sigmoid}((\log\mathbf{\alpha})/\beta))$, we add the term $\log \mathbf{u} - \log \mathbf{(1-u)}$ to $\log\mathbf{\alpha}$ before passing it through the Sigmoid function.

To perform weight pruning on a weight matrix $W\in\mathcal{R}^{d_1 \times d_2}$, we must first create a corresponding learnable mask $\mathbf{z}\in\mathcal{R}^{d_1\times d_2}$. Since we are using structured learning, we require two learnable masking parameters for this mask: $\mathbf{\alpha}_1\in\mathcal{R}^{d_1}$ and $\mathbf{\alpha}_2\in\mathcal{R}^{d_2}$. These parameters generate the input dimension mask $\mathbf{z}_1$ and the output dimension mask $\mathbf{z}_2$ with Eq.~\ref{eq:mask}, respectively.
We can then obtain a final mask $\mathbf{z}=\mathbf{z}_1\mathbf{z}_2^\top$, and use it to obtain a pruned weight matrix $W' = W \odot \mathbf{z}$, where $\odot$ represents the element-wise dot product.

In this paper, we set $\beta=1$, $\gamma=0$, and $\eta=1$, which implies that we do not stretch the output of the Sigmoid function. As a result, the hard-concrete distribution is equivalent to the concrete distribution~\cite{louizos2017learning}.

\section{Method}
\label{sec:method}
\subsection{Structured pruning FastSpeech 2}
\label{subsec:prune-fs2}
With the exception of the input and output dimensions, which are determined by the data, all dimensions in FastSpeech 2 are prunable. These dimensions are listed below:
\begin{itemize}
    \item The hidden dimension of the model $d$, which affects:
    \begin{itemize}
        \item The encoder/decoder's positional encoding.
        \item All embeddings' dimensions.
        \item Each MHA layer's $W_Q^{(i)},W_K^{(i)},W_V^{(i)},W_O^{(i)}$.
        \item Each FFN layer's $W_U,W_D$.
        \item Each layer-normalization layer's scale and shift.
        \item The input channels of the variance adaptor and the output linear layer.
    \end{itemize}
    \item Each MHA layer's $N_h$ and $d_k^{(i)}$, which affects $W_Q^{(i)},W_K^{(i)},W_V^{(i)},W_O^{(i)}$.
    \item Each FFN layer's $d_f^{(i)}$, which affects $W_U,W_D$.
    \item The hidden dimensions of the variance adaptor's and the post-net's hidden layers.
\end{itemize}
For each dimension mentioned above, we create a corresponding learnable masking parameter. For example, we use $\mathbf{\alpha}_d$ to mask the model dimension $d$, $\mathbf{\alpha}_k^{(i)}$ to mask MHA dimensions $d_k^{(i)}$, $\mathbf{\alpha}_f^{(i)}$ to mask FFN dimensions $d_f^{(i)}$, and $\mathbf{\alpha}_h$ to mask MHA heads $N_h$, etc.
During training, we generate a mask $\mathbf{z}$ for each TTS parameter based on its input/output connections, as illustrated in Fig.\ref{fig:prune}. For instance, since $d$ affects numerous parameters due to the residual connections in Eq.\ref{eq:transformer-block}, we must mask each of those parameters with a corresponding $\mathbf{z}$ based on $\mathbf{z}_d$, which is generated by the masking parameter $\mathbf{\alpha}_d$.


\subsection{Optimizing adaptive structured pruning masks}

To generate pruning masks $\mathbf{z}$ for all parameters in FastSpeech 2, we use the learnable masking parameters $\mathbf{\alpha}$, as described in Sec.\ref{subsec:l0} and Sec.\ref{subsec:prune-fs2}. We then compute the $L_1$ norm of all the masks and use this sum as the regularization term:
\begin{equation}
    \label{eq:reg}
    L_{reg} = \sum_\mathbf{z} \|\mathbf{z}\|_1.
\end{equation}
We initialize all $\mathbf{\alpha}$ with large values so that the sampled $\mathbf{z}$ would be all close to 1 at the start of training.
As we prune a voice-cloning model, the TTS loss becomes the loss term, resulting in the total loss as follows:
\begin{equation}
\begin{aligned}
    \label{eq:total}
    L_{total}=L_{TTS}+\frac{1}{\lambda} L_{reg}=L_{TTS}+\frac{1}{\lambda}\sum_\mathbf{z} \|\mathbf{z}\|_1,
\end{aligned}
\end{equation}
where $\lambda$ is a weighting factor for the regularization.
We set $\lambda$ to the total TTS parameters count in experiments, making the regularization term the model density (the portion of the model unpruned).


\subsection{Inference}

During inference, we skip using Eq.~\ref{eq:mask} to generate continuous pruning masks $\mathbf{z}$ from the hard-concrete distribution. Instead, we directly determine the binary pruning masks from each $\log\mathbf{\alpha}$.
As discussed in Sec.~\ref{subsec:l0}, the term $\mathrm{Sigmoid}((\log\mathbf{\alpha})/\beta)$ in Eq.~\ref{eq:mask} represents the probabilities of the Bernoulli distributions. We empirically observe that most of these probabilities are close to 0 or 1, while less than 2\% are within the range of (0.05, 0.95). Therefore, we use a threshold of $\mathrm{Sigmoid}((\log\mathbf{\alpha})/\beta)=0.5$ for simplicity. For each element $z_i$ in each $\mathbf{z}$ and its corresponding element $\mathbf{\alpha}_i$, we can calculate by the following condition:

\begin{equation}
    \label{eq:inference-mask}
    z_i=\left\{\begin{matrix}
    0, \;\;\;\mathrm{Sigmoid}((\log\mathbf{\alpha}_i)/\beta) < 0.5\\ 
    1, \;\;\;\mathrm{Sigmoid}((\log\mathbf{\alpha}_i)/\beta) \geq 0.5
\end{matrix}\right..
\end{equation}

\section{Experiments}
\label{sec:experiments}

\subsection{Setup}
We utilize \texttt{LibriTTS}~\cite{zen2019libritts} as our pre-training dataset and \texttt{VCTK}~\cite{yamagishi2019cstr} as our voice-cloning dataset. To transform the models' output Mel-spectrograms into waveforms, we use MelGAN~\cite{kumar2019melgan} as our vocoder.
The implementation of our FastSpeech 2 model and the training/pruning details can be found in our GitHub repository\footnote{\url{https://github.com/SungFeng-Huang/Meta-TTS}}.

In our experiments, we mainly focus on 8-shot voice cloning, where only 8 audio samples of the target speaker are used for fine-tuning and pruning.
For each speaker in \texttt{VCTK}, we randomly sample 8 recordings for a voice cloning task.
We pre-train the TTS models for 40k steps with \texttt{LibriTTS}, followed by fine-tuning/pruning the model with 8-shot voice cloning tasks until convergence.
The remaining recordings and their corresponding transcripts are utilized for evaluation. The transcripts are treated as testing inputs, and the recordings are considered as ground-truth baselines.

Through our experiments, we observe that the model dimension $d$ affects a significant number of parameters mentioned in Section~\ref{subsec:prune-fs2}, leading to a trade-off between TTS performance and model sparsity. To ensure the final voice cloning performance, we made the decision not to prune $d$ in our experiments.

We experiment with four settings: pruning before fine-tuning, after fine-tuning, pruning jointly with fine-tuning, and pruning with the pre-training data from \texttt{LibriTTS} before fine-tuning with the voice cloning data. The last setting is similar to the common pipeline of model distillation, where the model is first compressed and then fine-tuned.

\subsection{Subjective and objective evaluation metrics}
We assess the voice cloning performance based on the generated audio's quality and the speaker similarity. As \texttt{VCTK} comprises multiple accents, mostly unseen in \texttt{LibriTTS}, we also evaluate the synthesized speech's accent similarity to the target speaker.

For subjective evaluations, we randomly sample six testing inputs to synthesize for each task. We ask human raters to score the generated audio samples based on their \textbf{naturalness}, \textbf{speaker similarity}, and \textbf{accent similarity} using a 5-point (1-point increment) scale Mean Opinion Score (MOS). As reference, we provide two real recordings of the target speaker. The MOS test is conducted on Amazon Mechanical Turk (AMT), where each task is scored by at least five raters.

For our objective evaluations, we employ all testing inputs. We trained a speaker classifier using data from both \texttt{LibriTTS} and \texttt{VCTK}, which together comprise 2456 and 108 speakers, respectively. Additionally, we trained an accent classifier using the \texttt{VCTK} dataset, which features 12 distinct accents. Both classifiers utilize the x-vector~\cite{snyder2018x} as their model architecture and are trained with the SpecAugment++~\cite{wang2021specaugment++} data augmentation method to prevent overfitting. The speaker classifier attained a 97\% accuracy rate on the randomly-split test set (100\% accuracy on \texttt{VCTK} speakers), while the accent classifier achieved a 99\% accuracy rate on the randomly-split \texttt{VCTK} test set.

\subsection{Evaluation results and analysis}

\begin{table}[t]
  \caption{Subjective evaluations with standard deviations. High standard deviations are due to the varying scoring preference of the raters. ``$\xrightarrow{}$'' indicates the order of training stages.
  \textbf{GT}: ground-truth waveform. \textbf{GT + Vocoder}: ground-truth Mel-spectrogram with MelGAN vocoder. \textbf{FT}: fine-tune. \textbf{Prune}: prune with voice-cloning data. \textbf{Prune'}: prune with pre-training data (\texttt{LibriTTS}).}
  \label{tab:mos}
  \centering
  \scriptsize
  \begin{tabular}{llccc}
    \toprule
    \multirow{2}{*}{Approach} & \multirow{2}{*}{Stage} & \multicolumn{2}{c}{Similarity} & \multirow{2}{*}{Naturalness}\\
    \cmidrule(lr){3-4}
    & & Speaker & Accent &  \\
    \midrule
    \multicolumn{5}{l}{\textbf{Ground Truth}}\\
    \midrule
    GT &
        & $4.29_{(0.86)}$ & $4.21_{(0.90)}$ & $4.29_{(0.86)}$  \\
    GT + Vocoder &
        & $4.02_{(0.94)}$ & $3.96_{(0.94)}$ & $4.02_{(0.94)}$  \\
    \midrule
    \multicolumn{5}{l}{\textbf{Proposed}}\\
    \midrule
    Prune + FT & joint
        & $3.79_{(1.02)}$ & $3.73_{(1.01)}$ & $3.79_{(1.02)}$  \\
    \cmidrule(lr){1-5}
    \multirow{2}{*}{FT $\xrightarrow{}$ Prune}
        & 1st
        & ${3.83_{(1.05)}}$ & ${3.79_{(1.01)}}$  & ${3.83_{(1.05)}}$  \\
        & 2nd
        & ${3.81_{(1.04)}}$ & ${3.74_{(1.02)}}$  & ${3.81_{(1.04)}}$  \\
    \cmidrule(lr){1-5}
    \multirow{2}{*}{Prune $\xrightarrow{}$ FT}
        & 1st
        & ${3.77_{(1.05)}}$ & ${3.74_{(1.02)}}$  & ${3.77_{(1.05)}}$  \\
        & 2nd
        & ${3.77_{(1.04)}}$ & ${3.73_{(1.03)}}$  & ${3.77_{(1.04)}}$  \\
    \cmidrule(lr){1-5}
    \multirow{2}{*}{Prune' $\xrightarrow{}$ FT}
        & 1st
        & ${2.63_{(1.40)}}$ & ${2.86_{(1.27)}}$  & ${2.63_{(1.40)}}$  \\
        & 2nd
        & ${3.75_{(1.04)}}$ & ${3.69_{(1.05)}}$  & ${3.75_{(1.04)}}$  \\
    \bottomrule
  \end{tabular}
\end{table}

The results are shown in Table~\ref{tab:mos}.
The standard deviations of the scores are generally large, primarily due to the varying scoring preferences of the AMT raters. Although the majority of averaged scores are close to each other or have negligible differences, we observe that the ground truth recordings are rated significantly higher, while the model that only pruned by the pre-training data performs worse. Moreover, all the voice-cloned models receive high naturalness scores, indicating that their synthesized speech is of high quality. This observation is confirmed by performing t-tests over the MOS scores.


\begin{table}[t]
  \caption{Objective evaluations with standard deviations. \textbf{Sparsity} means the percentage of the parameters pruned. \textbf{Ratio} indicates the compression ratio of the model (how much smaller).}
  \label{tab:objective}
  \centering
  \scriptsize
  \begin{tabular}{llcccc}
    \toprule
    \multirow{2}{*}{Approach} & \multirow{2}{*}{Stage} & \multirow{2}{*}{Sparsity (\%)} & \multirow{2}{*}{Ratio} & \multicolumn{2}{c}{Accuracy} \\
    \cmidrule(lr){5-6}
    & & & & Speaker & Accent  \\
    \midrule
    Prune + FT  & joint & $\mathbf{85.9}_{(1.62)}$   & $\mathbf{7.1\times}$  & $0.960_{(0.130)}$ & $0.941_{(0.190)}$   \\
    \cmidrule(lr){1-6}
    \multirow{2}{*}{FT $\xrightarrow{}$ Prune}
        & 1st   & $0.00_{(0.00)}$   & $-$ & ${0.912_{(0.207)}}$ & ${0.961_{(0.149)}}$    \\
        & 2nd   & $81.8_{(1.74)}$   & $5.5\times$   & ${0.959_{(0.101)}}$ & ${0.993_{(0.030)}}$    \\
    \cmidrule(lr){1-6}
    \multirow{2}{*}{Prune $\xrightarrow{}$ FT}
        & 1st   & $83.2_{(0.76)}$   & $6.0\times$   & ${0.747_{(0.266)}}$ & $0.972_{(0.078)}$   \\
        & 2nd   & $83.2_{(0.76)}$   & $6.0\times$   & $\mathbf{0.965}_{(0.035)}$ & $\mathbf{0.996}_{(0.014)}$   \\
    \cmidrule(lr){1-6}
    \multirow{2}{*}{Prune' $\xrightarrow{}$ FT}
        & 1st   & $76.6_{(0.40)}$   & $4.3\times$   & ${0.000_{(0.000)}}$ & ${0.218_{(0.179)}}$   \\
        & 2nd   & $76.6_{(0.40)}$   & $4.3\times$   & ${0.928_{(0.130)}}$ & ${0.980_{(0.089)}}$   \\
    \bottomrule
  \end{tabular}
\end{table}

Table~\ref{tab:objective} presents the objective results. Our speaker classifier serves as a rigorous evaluation metric for objective evaluations.
Since we require the speaker classifier to identify samples from among more than 2.5k speakers, the audio samples must be extremely similar to the target speaker for the classifier to predict correctly. Otherwise, the speaker classifier may easily misclassify the samples as belonging to other speakers.
Surprisingly, all fine-tuned models performed exceptionally well, exhibiting high speaker and accent accuracies. However, the voice-cloned model without compression did not achieve the best performance.
Pruning followed by fine-tuning produced the highest speaker and accent accuracies, with the slightest standard deviations and the second-largest sparsity.
Hence, we assert that pruning before fine-tuning is a robust and stable training pipeline for voice cloning.
Intriguingly, even when we only prune the model without fine-tuning, it still achieves a 74.7\% speaker accuracy and 97.2\% accent accuracy.
This indicates that, even if the TTS model has never encountered the target speaker during pre-training, it may still contain a sub-network capable of achieving high speaker and accent accuracy.

Despite using different training pipelines, all fine-tuned models yield comparable speaker and accent accuracy, but not in terms of model compression ratio.
Joint pruning and fine-tuning yields the highest level of compression (85.9\% sparsity, $7.1\times$ smaller) among all training pipelines.
We hypothesize that the model learns to remove unnecessary components and optimize unpruned parameters through joint pruning and fine-tuning. In contrast, pruning before fine-tuning compresses the model by a factor of 6.0 (83.2\% sparsity), while pruning after fine-tuning compresses the model by a factor of 5.5 (81.8\% sparsity).
Pruning with pre-training data before fine-tuning yields the worst compression ratio (76.6\% sparsity), possibly because the pre-training data forces the model to maintain its high-quality audio generation capability for all pre-training speakers. However, when pruning with voice cloning data, the model only needs to develop its generation capacity for a single target speaker, making it easier and requiring fewer parameters.

\subsection{Other pruning advantages}
The pruned model can double the inference speed and cut peak GPU usage in half. Additionally, unlike model distillation, another architecture reduction method, model pruning does not necessitate training from scratch, significantly reducing training time. Moreover, model distillation struggles to achieve good audio quality for training the small TTS models from scratch, whereas model pruning methods initialize TTS models from high-quality pre-trained TTS models and maintain audio quality throughout the pruning process.

\section{Conclusion}

We propose using speaker-adaptive structured pruning for voice cloning to create personalized TTS models that are as lightweight as possible, making them more suitable for deployment on mobile devices. In our experiments, we compared different voice-cloning training pipelines and discovered that pruning before fine-tuning is the most stable pipeline for obtaining a compressed voice cloning model with high speaker and accent accuracies. However, jointly pruning with fine-tuning yields the most compressed voice cloning model with a size of $7.1\times$ smaller than the original TTS model, with comparable performance. In summary, applying model pruning to voice cloning reduces model size and achieves comparable or even better voice cloning performance.

\vfill\pagebreak
\section{Acknowledgement}
We thank to National Center for High-performance Computing (NCHC) for providing computational and storage resources.

\bibliographystyle{IEEEbib}
\bibliography{refs}

\end{document}